\documentclass[aps,prl,showpacs,twocolumn,a4paper,superscriptaddress,10pt]{revtex4-2}
\usepackage{etoolbox}
\apptocmd{\sloppy}{\hbadness 2000\relax}{}{}

\usepackage[english]{babel}
\usepackage{graphicx}
\usepackage{hyperref}
\hypersetup{
    colorlinks,
    linkcolor={red!50!black},
    citecolor={blue!50!black},
    urlcolor={blue!80!black}}
\usepackage{cancel}
\usepackage{amsmath}
\usepackage{amsfonts}
\usepackage{amssymb}
\usepackage{bm}

\usepackage{physics}
\usepackage{relsize}
\usepackage{mathtools}
\usepackage{indentfirst}
\usepackage{pstricks}
\usepackage{tikz}
\tikzset{>=latex}
\usetikzlibrary{patterns}
\usepackage[overload]{empheq}
\usepackage{xcolor}
\usepackage{natbib}
\usepackage{dsfont}
\usepackage{microtype}
\usepackage{mathrsfs}

\usepackage[caption=false]{subfig}

\makeatletter
\renewcommand{\fnum@figure}{FIG. \thefigure}
\makeatother

\definecolor{Mblue}{rgb}{0.368417, 0.506779, 0.709798}
\definecolor{Myel}{rgb}{0.880722, 0.611041, 0.142051}
\definecolor{Mgre}{rgb}{0.560181, 0.691569, 0.194885}

\begin{document}
\title{Generic Stress Rectification in Nonlinear Elastic Media}
\author{F\'elix Benoist}
\affiliation{Universit\'e Paris-Saclay, CNRS, LPTMS, 91400, Orsay, France}
\email{felix.benoist@lptms.u-psud.fr}
\author{Guglielmo Saggiorato}
\affiliation{Universit\'e Paris-Saclay, CNRS, LPTMS, 91400, Orsay, France}
\author{Martin Lenz}
\email{martin.lenz@universite-paris-saclay.fr}
\affiliation{Universit\'e Paris-Saclay, CNRS, LPTMS, 91400, Orsay, France}
\affiliation{PMMH, CNRS, ESPCI Paris, PSL University, Sorbonne Universit\'e,
Universit\'e de Paris, F-75005, Paris, France}

\begin{abstract}
Stress propagation in nonlinear media is crucial in cell biology, where molecular motors exert anisotropic force dipoles on the fibrous cytoskeleton. While the force dipoles can be either contractile or expansile, a medium made of fibers which buckle under compression rectifies these stresses towards a biologically crucial contraction. A general understanding of this rectification phenomenon as a function of the medium's elasticity is however lacking. Here we use theoretical continuum elasticity to show that rectification is actually a very general effect in nonlinear materials subjected to anisotropic internal stresses. We analytically show that both bucklable and constitutively linear materials subjected to geometrical nonlinearities rectify small forces towards contraction, while granular-like materials rectify towards expansion. Using simulations, we moreover show that these results extend to larger forces. Beyond fiber networks, these results could shed light on the propagation of stresses in brittle or granular materials following a local plastic rearrangement.
\end{abstract}


\maketitle

The active, stress-generating role of many biological materials stems from their ability to transmit internally generated forces. In cells, the action of molecular motors and the growth of protein fibers over a few nanometers generates anisotropic forces that are further transmitted by a fibrous network, the cytoskeleton, to the scale of the whole cell~\cite{howard_mechanics_2001, blanchoin_actin_2014}. At larger length scales, polarized cells in connective tissues exert anisotropic stresses on another fibrous network, the extracellular matrix, which again propagates these stresses far from their application point~\cite{Schwarz:2013, maskarinec_quantifying_2009}.

The well-characterized nonlinear stress response of these networks~\cite{gardel_elastic_2004, storm_nonlinear_2005, chaudhuri_reversible_2007} plays a crucial role in force transmission, allowing for the enhancement of contractile stresses~\cite{Murrell12, ronceray_fiber_2016, Ronceray_2019,Han:2018} and promoting long-range mechano-sensitivity~\cite{rosakis_model_2015, notbohm_microbuckling_2015, xu_nonlinearities_2015, wang_long-range_2014, Sopher:2018}. Beyond this quantitative stress amplification, the nonlinear response of fiber networks also leads to qualitative changes in the propagated stresses, as previously shown in numerical simulations~\cite{ronceray_fiber_2016}. In these simulations, a localized active unit exerts anisotropic forces in the center of a large network of discrete fibers, each of which can buckle under a sufficiently large compressive force. For localized forces much larger than this buckling threshold, the far-field stresses transmitted by the network become contractile. This is valid even in cases where the local forces are predominantly expansile, because the network resists and therefore propagates tension more than compression. This stress ``rectification'' has strong implications for biological force propagation, and could be one of the reasons why the actomyosin cytoskeleton is overwhelmingly observed to contract irrespective of its detailed internal architecture.

Here, we generalize these results beyond bucklable fiber networks, and demonstrate that stress rectification is a generic corollary of stress propagation in a nonlinear elastic medium. Our approach is based on a continuum formalism that allows a general discussion of arbitrary nonlinearities. We consider both geometrical nonlinearities and generic material-dependent nonlinearities describing the response of the material to compression or tension. Nonlinearities whereby the material stiffens under tension and soften under compression are characteristic of bucklable fiber networks~\cite{storm_nonlinear_2005}. Conversely, materials that soften under tension and stiffen under compression, or ``anti-buckle'', may offer a description of granular media, where contacts between grains are disrupted as the confining pressure is decreased~\cite{Ellenbroek_PRL09}. Under shear, these materials experience localized plastic events known as shear transformations which generate anisotropic internal stresses similar to those induced by molecular motors in the cytoskeleton~\cite{Amon12}. We show that the elastic constants describing the weakly nonlinear response of these materials are a reliable predictor of the sign and magnitude of rectification.

We consider a piece of homogeneous, isotropic elastic medium of dimension $d$ comprised in a domain $\Omega$. A set of anisotropic ``active units'' (\emph{e.g.}, molecular motors or shear transformation zones) exerts forces and/or imposes local displacements on the medium. This induces a force density $\mathbf{f}$, resulting in a Cauchy stress tensor $\bm\sigma$ given by the force balance equation $f_i=-\partial\sigma_{ij}/\partial X_j$. Here $\bf X=x+u$ is the final location (in the ``target space'') of a material point initially located in $\mathbf{x}$ (in the ``initial space''), $\bf u$ denotes the displacement vector and the summation over repeated indices is implied. The boundary $\partial\Omega$ of the medium is held fixed, such that the forces exerted by the active units are transmitted through the medium and cause it to exert a coarse-grained stress 
\begin{equation}\label{eq:active_stress}
	\bar{\sigma}^a_{ij}=\frac{1}{V}\oint_{\partial\Omega}\sigma_{ik}X_j \,\textrm{d}A_k
\end{equation}
onto the boundary~\cite{ronceray_connecting_2015}, where $V$ is the volume of the medium and $\text{d}\mathbf{A}$ the outward-directed area element in the target space. In the context of active matter, $\bar{\bm\sigma}^a$ is known as the active stress generated by the overall system comprised by the medium and the active units~\cite{Prost:2015}. We define as contraction (expansion) a situation where the active pressure $P_a=-\bar\sigma_{ii}^a/d$ is negative (positive). To investigate the relationship between the local forces $\mathbf{f}$ and the active stress $\bar{\bm\sigma}^a$, we define the coarse-grained local stress
\begin{equation}\label{eq:local_stress}
	\bar{\sigma}^l_{ij}=-\frac{1}{V}\int_\Omega f_iX_j\,\text{d}V,
\end{equation}
where $\text{d}V$ is the volume element in the target space. In the special case where the force transmission is entirely linear, this relation simply reads $\bar{\bm\sigma}^a=\bar{\bm\sigma}^l$, implying in particular an equality of the active and local pressures $P_a=P_l=-\bar\sigma_{ii}^l/d$.  In that case, contractile (expansile) local forces always imply a contractile (expansile) active stress. These equalities are however violated in nonlinear media~\cite{ronceray_connecting_2015, carlsson_contractile_2006}, and the local and active pressures $P_l$ and $P_a$ can have opposite signs. We show here that this stress rectification may arise from geometrical and/or constitutive nonlinearities in the material's elastic response, and that geometrical nonlinearities always bias the system towards contraction. We then investigate the effect of generic, lowest-order constitutive nonlinearities, and characterize the regimes conducive to rectification towards contraction and expansion. Finally, we use finite-element simulations to show that our conclusions remain qualitatively valid at higher orders. 

We describe the elastic deformation of our medium using the displacement gradient $\eta_{ij}=\partial u_i/\partial x_j$ and introduce the Green-Lagrange strain tensor $\bm\varepsilon=(\bm\eta+\bm\eta^\mathsf T+\bm\eta^\mathsf T\bm\eta)/2$~\cite{Wriggers08}. The last, nonlinear term of $\bm\varepsilon$ is purely geometrical and accounts for, \emph{e.g.}, material rotations. We express the Cauchy stress as a function of the elastic energy density $E$ in the initial space by $\bm\sigma=(\mathbf1+\bm\eta)\frac{\partial E}{\partial\bm\varepsilon}(\mathbf1+\bm\eta^\mathsf T)/\det(\mathbf1+\bm\eta)$, where $\mathbf{1}$ denotes the unit tensor. We first consider a constitutively linear material with a quadratic energy density $E=\kappa\varepsilon_{ii}^2/2+\mu(\varepsilon_{ij}^2-\epsilon_{ii}^2/d)$, where $\kappa$ and $\mu$ are the bulk and shear moduli. We use the divergence theorem to turn the right-hand side of Eq.~\eqref{eq:active_stress} into a volume integral, and combine the expression of the Cauchy stress, the force balance equation and Eq.~\eqref{eq:local_stress} to find 
\begin{equation}\label{eq:linear}
\begin{aligned}
	P_a&= P_l -\int_\Omega\frac{\text{d}v}{Vd}\bigg[\frac{\kappa}{2}\left(d\eta_{ij}^2+4\varepsilon_{ii}^2\right)+4\mu\big(\varepsilon_{ij}^2-\varepsilon_{ii}^2/d\big)\bigg]\le P_l,\hspace{-.6cm}
\end{aligned}
\end{equation}
where the integral runs over the initial space. The inequality in Eq.~\eqref{eq:linear} is proven in the SI and means that the system as a whole is always more contractile than the local forces, implying that geometrical nonlinearities always induce a rectification towards contraction.

To describe nonlinearities resulting from the medium's constitutive properties, we consider a two-dimensional isotropic, achiral elastic medium with a non-harmonic energy density:
\begin{equation}\label{eq:nonlin_energy}
	E=\frac{\kappa+\kappa'\varepsilon_{ii}/3}{2}\varepsilon_{ii}^2+\frac{\mu+\mu'\varepsilon_{ii}}{d}\big(d\varepsilon_{ij}^2-\varepsilon_{ii}^2\big) +\order{\eta^4},\hspace{-.09cm}
\end{equation}
where the coefficients $\kappa'$, $\mu'$ can be of either sign and characterize the most general, lowest-order nonlinearity. According to Eq.~\eqref{eq:nonlin_energy}, when the material is isotropically dilated by a relative amount $\varepsilon_{ii}\sim\delta V/V_0$ its bulk (shear) modulus exceeds 
that of a purely harmonic material by $\kappa' \delta V/V_0$ ($\mu' \delta V/V_0$). More generally, we may consider a combination of bulk expansion and simple shear
\begin{equation}
	\bm\eta = \begin{pmatrix}\eta_{ii}/2&\eta_{xy}\\0&\eta_{ii}/2\end{pmatrix},
\end{equation}
compute the Cauchy stress tensor, and derive the differential bulk and shear moduli as 
\begin{subequations}\label{eq:moduli}
\begin{align}
	K & =\frac{\partial\sigma_{xx}}{\partial\eta_{ii}}
		=\kappa\left(1+\kappa_1\eta_{ii}\right)+\order{\eta^2},\\
	G & =\frac{\partial\sigma_{xy}}{\partial\eta_{xy}}
		=\mu\left(1+\mu_1\eta_{ii}\right)+\order{\eta^2},
\end{align}
\end{subequations}
where the first order nonlinear corrections to the moduli ${\kappa}_1=1/2+\kappa'/\kappa$ and ${\mu}_1=\kappa/\mu+1/2+\mu'/\mu$ include contributions from geometrical as well as constitutive nonlinearities. 
Based on rheology measurements, we estimate $\kappa_1\approx 100$ and $\mu_1\approx 30$ for gels of the extracellular matrix filaments fibrin and collagen~\cite{vanOosten2016}. These positive values are consistent with the notion that biological fiber networks buckle, and therefore soften, under compression ($\eta_{ii}<0$). Conversely, granular materials tend to increase their cohesion under compression. Experiments and simulations on polydisperse soft spheres near jamming thus suggest $\kappa_1\approx0$ and $\mu_1 \in [-400,-4]$~(see Refs.~\cite{O'Hern03,vanHecke09} and SI). An intermediate behavior is observed in fiber networks with stiff grain-like inclusions mimicking connective tissues. This gives rise to a more complicated sign combination which depends on the inclusion density~\cite{Shivers2020, VanOosten2019}. Finally, a standard (``neo-Hookean'') model of rubber displays $\kappa_1>0$ and $\mu_1<0$ with small values~\cite{treloar_elasticity_1973, shokef_scaling_2012}, see SI.

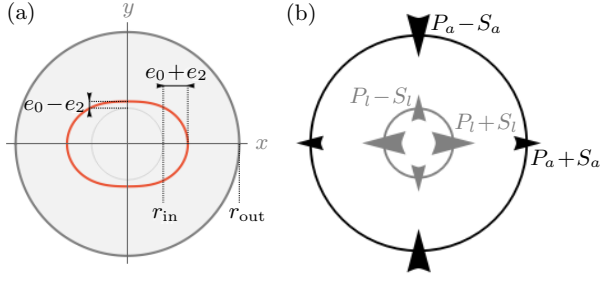
\begin{figure}[!t]
    \centering
    \begin{tikzpicture}
    	\node[anchor=north west,inner sep=0pt,outer sep=0pt] at (0,0) 
   		{\includegraphics[width=.21\textwidth]{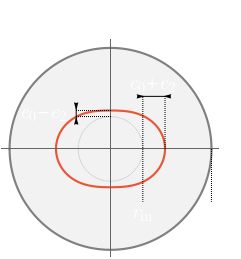}  };
    	\draw (-.1,-.48) node[right] {(a)};
    	\draw[white] (-.1,-3.92) node[right] {(b)};
    	\draw[gray] (1.54,-.43) node[right] {$y$};
    	\draw (.14,-1.76) node[right] {$e_0\!-\!e_2$};
    	\draw (1.86,-1.31) node[right] {$e_0\!+\!e_2$};
    	\draw[gray] (3.42,-2.33) node[right] {$x$};
    	\draw (1.96,-3.32) node[right] {$r_\text{in}$\hspace{.7cm}$r_\text{out}$};
    \end{tikzpicture}   \hspace{-.2cm} 
	\begin{tikzpicture}
    	\node[anchor=north west,inner sep=0pt,outer sep=0pt] at (0,0) 
   		{\includegraphics[width=.21\textwidth]{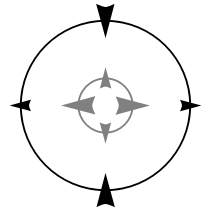}  };
    	\draw (-.1,-.06) node[right] {(b)};
    	\draw (1.94,-.2) node[right] {$P_a\!-\!S_a$};
    	\draw[gray] (.8,-1.23) node[right] {$P_l\!-\!S_l$};
    	\draw[gray] (2.28,-1.6) node[right] {$P_l\!+\!S_l$};
    	\draw (3.35,-2.1) node[right] {$P_a\!+\!S_a$};
    \end{tikzpicture}
   \caption{Sketches of the imposed anisotropic displacement and the resulting coarse-grained stresses. (a)~In the target configuration, the inner light-gray circle with radius $r_\text{in}$ is moved to the orange ring. (b)~Stress components in a particular situation where the local pressure $P_l$ is positive and the active pressure $P_a$ at the boundary is negative.}
\label{DiskSchematic}
\end{figure}

To explicitly predict the active pressure resulting from rectification, we consider a simple circular piece of elastic medium with radius $r_\text{out}$ and a single active unit at its center. The active unit is a circle with radius $r_\text{in}$ at rest, and undergoes a radial displacement [Fig.~\ref{DiskSchematic}(a)]
\begin{equation}\label{eq:u_l}
	\mathbf u(r_\text{in}) = r_\text{in}\left[e_0+e_2\cos(2\theta)\right]\hat{\mathbf r}.
\end{equation}
This induces a mixture of compression, tension and shear on the medium. Symmetry imposes that the local and active stress tensors take the form
\begin{equation}
	\bar{\bm\sigma}^x = -\mathsmaller{\begin{pmatrix}P_x+S_x&0\\0&P_x-S_x\end{pmatrix}},
\end{equation}
in Cartesian coordinates, for $x\in\lbrace l,a\rbrace$ [Fig.~\ref{DiskSchematic}(b)].
As shown in Eq.~\eqref{eq:local_stress}, the local coarse-grained stress $\bar{\bm\sigma}^l$ is the ratio of a force dipole by the volume $V$. Assuming a constant local dipole, $\bar{\bm\sigma}^l$ thus decreases with increasing system size $V$ due to dilution. A similar statement holds for $\bar{\bm\sigma}^a$. It is thus useful for our discussion to define the quantities $\mathcal P_x=P_x(r_\text{out}/r_\text{in})^2$ and $\mathcal S_x=S_x(r_\text{out}/r_\text{in})^2$ which are not subject to this dilution. In this sense, they behave as force dipole components. In the following, we consider the lowest order in the weakly nonlinear regime $e_0, e_2\ll 1$ (see SI). We perturbatively solve the force balance equation using Eq.~\eqref{eq:u_l} as well as the fixed boundary condition in $r_\text{out}$ to compute the pressure and shear components $\mathcal P_x$, $\mathcal S_x$ as
\begin{subequations}\label{eq:P_x}
\begin{align}
	\mathcal P_x &= A_xe_0+ B_xe_2^2 + \order{e_0^2,e_0e_2^2,e_2^4},\\
	\mathcal S_x &= C_x e_2 + \order{e_2e_0,e_2^3},
\end{align}
\end{subequations}
where the cumbersome dependence of $A_x$, $B_x$ and $C_x$ on the properties of the medium is detailed in the SI. The active stresses can then be computed from the local ones through
\begin{equation}\label{eq:Pa=f(Pl)}
	\mathcal P_a\sim \mathcal P_l+\alpha \mathcal S_l^2,\quad \mathcal S_a\sim \mathcal S_l.
\end{equation}
Here $\alpha \mu$ is a dimensionless function of $r_\text{out}/r_\text{in}$, $\kappa/\mu$, $\kappa_1$ and $\mu_1$ that is obtained by combining $A_x$, $B_x$ and $C_x$. At this order in nonlinearity, stress propagation in a medium with $\alpha =0$ resembles that in a linear medium (namely $\mathcal P_a=\mathcal P_l$, $\mathcal S_a=\mathcal S_l$). Conversely, a medium with a negative (positive) $\alpha$ harnesses the anisotropy of the active unit to produce an additional medium-wide contraction (expansion). Equation~\eqref{eq:Pa=f(Pl)} is formally valid for local stresses much smaller than the elastic moduli of the medium ($\mathcal P_l,\mathcal S_l\ll \kappa$, where ``$\kappa$'' stands for the typical magnitude of the linear moduli). It implies that when $\sqrt{\kappa \mathcal P_l}\ll \mathcal S_l$, the sign of the active pressure induced by a highly anisotropic active unit is determined not by the values $(\mathcal P_l, \mathcal S_l)$ characterizing the active unit, but by the properties of the medium through the sign of $\alpha$.

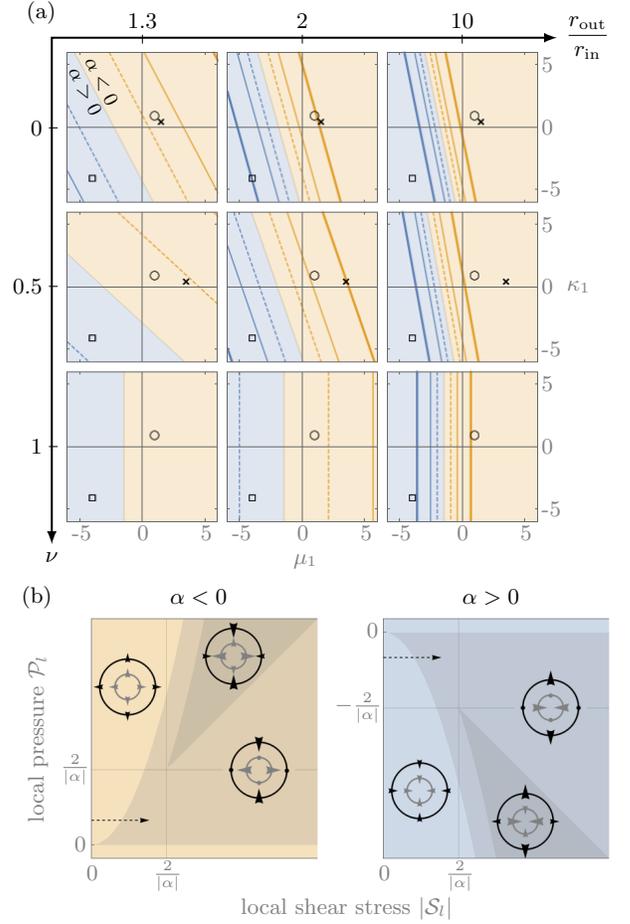
\begin{figure}[!t]
    \centering
	\begin{tikzpicture}[scale=1.18]
    	\node[anchor=north west,xshift=+1mm,yshift=-.75mm] at (0,0)
   		{\includegraphics[width=.77\linewidth]{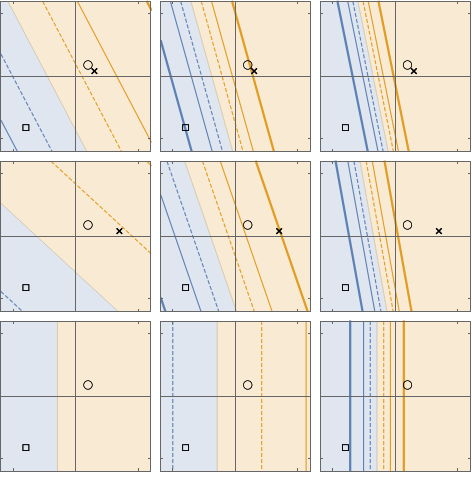}};
   		\draw (-.4,.3) node[right] {(a)};
    	\draw [<->,thick] (0,-6.07) node (yaxis) [below] {$\nu$} |- (6.05,0) node (xaxis) [right] {$\dfrac{r_\text{out}}{r_\text{in}}$};
	    \draw(1.085,1pt)--(1.085,-2pt) node[anchor=south,yshift=+1mm]{1.3};				\draw(3.00,1pt)--(3.00,-2pt) node[anchor=south,yshift=+1mm]{2};					\draw(4.915,1pt)--(4.915,-2pt) node[anchor=south,yshift=+1mm]{10};				\draw(-1pt,-1.08)--(2pt,-1.08) node[anchor=east,xshift=-1mm]{0}; 				\draw(-1pt,-2.98)--(2pt,-2.98) node[anchor=east,xshift=-1mm]{0.5};				\draw(-1pt,-4.9)--(2pt,-4.9) node[anchor=east,xshift=-1mm]{1};					\draw(.58,-1.0) node[anchor=east,rotate=-63.5]{\scalebox{.9}{$\alpha>0$}};
    	\draw(.775,-.885) node[anchor=east,rotate=-63.5]{\scalebox{.9}{$\alpha<0$}};
    	\foreach \x in {0,1.915,3.83}
    		{\draw[gray] (5.77,-\x-.32) node[right]{5};
    		\draw[gray] (5.77,-\x-1.07) node[right]{0};
    		\draw[gray] (5.77,-\x-1.815) node[right]{-5};}
    	\draw[gray] (6.07,-1.915-1.07) node[right]{$\kappa_1$};
    	\foreach \x in {0,1.915,3.83}
    		{\draw[gray] (\x+.145,-5.94) node[right]{-5};
    		\draw[gray] (\x+.925,-5.94) node[right]{0};
    		\draw[gray] (\x+1.685,-5.94) node[right]{5};}
   		\draw[gray] (1.915+.88,-6.27) node[right]{$\mu_1$};
	\end{tikzpicture}
	\begin{tikzpicture}
   		\node[anchor=north west,inner sep=0pt,outer sep=0pt] at (0,0) 
		{\includegraphics[width=.37\linewidth]{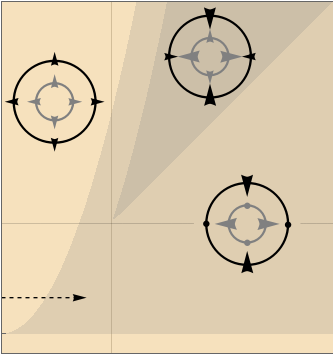}  };
		\draw (-1.1,.3) node[right] {(b)};
		\draw (1.,.3) node[right] {$\alpha<0$};
    	\draw[gray] (-.55,-2.14) node[right]{$\frac{2}{\abs{\alpha}}$};
    	\draw[gray] (-.35,-3.2) node[right]{0};
    	\draw[gray] (-.75,-3.) node[right,rotate=90] {local pressure $\mathcal P_l$};
    	\draw[gray] (-.18,-3.6) node[right]{0};
    	\draw[gray] (.76,-3.65) node[right]{$\frac{2}{\abs{\alpha}}$};
   		\draw[gray] (2.,-4.1) node[right] {local shear stress $|\mathcal S_l|$};
    \end{tikzpicture}\hspace{-2.cm}
    \begin{tikzpicture}
   		\node[anchor=north west,inner sep=0pt,outer sep=0pt] at (0,0) 
		{\includegraphics[width=.37\linewidth]{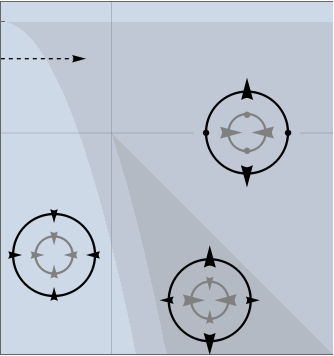}  };
		\draw (1.,.3) node[right] {$\alpha>0$};
    	\draw[gray] (-.35,-.18) node[right]{0};
    	\draw[gray] (-.8,-1.27) node[right]{$-\frac{2}{|\alpha|}$};
    	\draw[gray] (-.18,-3.6) node[right]{0};
    	\draw[gray] (.76,-3.65) node[right]{$\frac{2}{|\alpha|}$};
   		\draw[white] (1.3,-4.1) node[right] {$|\mathcal S_l|$};
    \end{tikzpicture}
    \caption{Bucklable materials ($\kappa_1,\mu_1>0$) rectify towards contraction (yellow), while very anti-bucklable materials ($\kappa_1,\mu_1<-3/2$) rectify towards expansion (blue). (a)~Contour plot of $\alpha$, indicating the overall sign of rectification as a function of the relative system size $r_\text{out}/r_\text{in}$, the Poisson ratio $\nu$ ($\nu=1$ denotes incompressibility in 2D) and the nonlinear corrections to the moduli ${\kappa}_1$, ${\mu}_1$. The contour lines denote $|\alpha|\mu$ = 2 (thick), 1 (thin), 0.5 (dashed). Crosses indicate constitutively linear materials where only geometrical nonlinearities are present (for $\nu=1$ they are far to the right). Circles and squares point out specific media discussed in Fig.~\ref{FES_Shokef}. (b)~Dependence of the signs of the components of the active stress (dark arrows in the insets) as functions of the local stress components. Regions without shading correspond to situations where the signs are the same as in the absence of rectification. In regions with intermediate shading ($|\mathcal P_l|\lesssim|\alpha|\mathcal S_l^2$), the sign of $\mathcal P_a$ is reversed. In the dark regions, $|\mathcal P_l|$ and $|\mathcal S_l|$ are so large that all components of $\bar{\sigma}_a$ (dark regions) are reversed~[SI]. These changes of signs are illustrated by arrows in the small pictures. Some arrows are replaced by circles in the intermediate shading regime to indicate that they are smaller than the other arrows and can point either way.}
\label{K1M1_full}
\end{figure}

We illustrate the influence of the material's properties on the sign of $\alpha$ in Fig.~\ref{K1M1_full}(a), which indicates a clear tendency of fiber-like (granular-like) materials towards contractile (expansile) rectification. Indeed, when ${\kappa}_1$ and ${\mu}_1$ are both larger (smaller) than a critical value of $-3/2$, the system always rectifies towards contraction (expansion). As a result, a material with  ${\kappa}_1={\mu}_1=0$ is contractile because of the contractile character of geometrical nonlinearities described by Eq.~\eqref{eq:linear}. Media with ${\kappa}_1>-3/2$ but ${\mu}_1<-3/2$ or the reverse can be either contractile or expansile depending on the system size $r_\text{out}/r_\text{in}$ and Poisson's ratio $\nu=(\kappa-\mu)/(\kappa+\mu)$. Finally, $|\alpha|$ increases with increasing $r_\text{out}$ such that $|\alpha(\infty)-\alpha(r_\text{out})|\propto (r_\text{in}/r_\text{out})^{2}$ for large $r_\text{out}$ (see SI), implying that larger systems rectify more. For example, larger fiber networks allow for more extensive buckling, resulting in stronger rectification and the coming together of the contour lines of Fig.~\ref{K1M1_full}(a) as $r_\text{out}$ increases. Finally, Fig.~\ref{K1M1_full}(b) shows that for large enough local stresses, rectification can cause a sign-switching not only in the active pressure but in all components of the active stress tensor $\bar{\sigma}_a$.

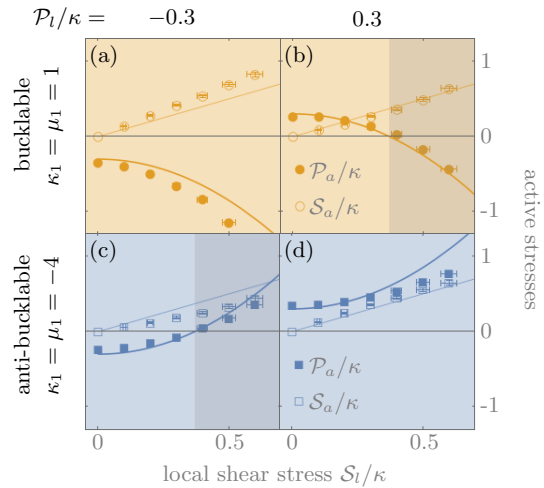
\begin{figure}[!t]
\hspace{-.3cm}\begin{minipage}[b]{1.\linewidth}
	\subfloat {\begin{tikzpicture}  
   		\node[anchor=north west,inner sep=0pt,outer sep=0pt] at (0,0) 
		{\includegraphics[width=.32\linewidth]{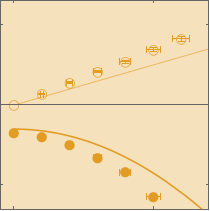}  };
    	\draw (-.06,-.25) node[right] {(a)};
    	\draw (-.85,.285) node[right] {$\mathcal P_l/\kappa=$\hspace{.6cm}$-0.3$};
    	\draw (-.9,-2) node[right, rotate=90] {bucklable};
    	\draw (-.45,-2.2) node[right, rotate=90] {$\kappa_1=\mu_1=1$};
    \end{tikzpicture}\hspace{-.19cm} 
	\begin{tikzpicture} 
    	\node[anchor=north west,inner sep=0pt,outer sep=0pt] at (0,0) 
   		{\includegraphics[width=.32\linewidth]{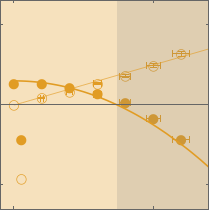}  };
    	\draw (-.06,-.25) node[right] {(b)};
    	\draw (.91,.285) node[right] {$0.3$};
    	\draw[gray] (.3,-1.89) node[right] {$\mathcal P_a/\kappa$};
    	\draw[gray] (.3,-2.4) node[right] {$\mathcal S_a/\kappa$};
    	\draw[gray] (2.72,-.31) node[right]{1};
    	\draw[gray] (2.72,-1.37) node[right]{0};
    	\draw[gray] (2.72,-2.42) node[right]{-1};
	\end{tikzpicture}  
		}  \vspace{-.425cm}\\
	\subfloat {\hspace{-.11cm}\begin{tikzpicture}
   		\node[anchor=north west,inner sep=0pt,outer sep=0pt] at (0,0) 
   		{\includegraphics[width=.32\linewidth]{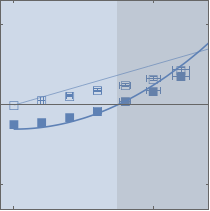}  };
   		\draw (-.06,-.25) node[right] {(c)};
   		\draw (-.9,-2.4) node[right, rotate=90] {anti-bucklable};
   		\draw (-.45,-2.4) node[right, rotate=90] {$\kappa_1=\mu_1=-4$};
   		\draw[gray] (-.02,-2.94) node[right]{0};
   		\draw[gray] (1.72,-2.94) node[right]{0.5};
   		\draw[gray] (.98,-3.41) node[right]{local shear stress $\mathcal S_l/\kappa$};
   		\draw (.78,-3.3) node[right] {$ $};
   	\end{tikzpicture}\hspace{-1.79cm}
    \begin{tikzpicture}
   		\node[anchor=north west,inner sep=0pt,outer sep=0pt] at (0,0) 
   		{\includegraphics[width=.32\linewidth]{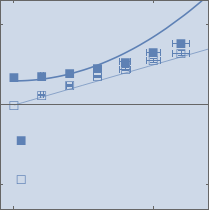}  };
   		\draw (-.06,-.25) node[right] {(d)};
    	\draw[gray] (.3,-1.89) node[right] {$\mathcal P_a/\kappa$};
    	\draw[gray] (.3,-2.4) node[right] {$\mathcal S_a/\kappa$};
    	\draw[gray] (2.72,-.31) node[right]{1};
    	\draw[gray] (2.72,-1.37) node[right]{0};
    	\draw[gray] (2.72,-2.42) node[right]{-1};
   		\draw[gray] (-.02,-2.94) node[right]{0};
   		\draw[gray] (1.72,-2.94) node[right]{0.5};
   		\draw[gray] (1.78,-3.44) node[right]{\color{white}$\mathcal S_l$};
   		\draw (.78,-3.3) node[right] {$ $};
    \end{tikzpicture}}
\end{minipage}\hspace{-.8cm}
	\begin{tikzpicture}
		\draw (0,0) node[right] {$ $};
		\draw[gray] (0,4.6) node[right, rotate=-90] {active stresses};
	\end{tikzpicture}
   \caption{The small-stress asymptotic prediction of Eq.~\eqref{eq:Pa=f(Pl)} (lines) accurately capture the finite-element simulation results (symbols) even for intermediate stress values. Here $\nu=0.1$ and $r_\text{out}/r_\text{in}=2$ in the geometry of Fig.~\ref{DiskSchematic}. (a,b)~A fiber-like bucklable model, (c,d)~a very anti-bucklable model mimicking a granular medium. The values of $\mathcal P_l$, $\mathcal S_l$ pictured in (b,c) are marked by dashed arrows in Fig.~\ref{K1M1_full}(b), and background shading follows the same convention. The error bars denote the estimated magnitude of the error induced by the finiteness of the simulation mesh size.}
\label{FES_Shokef}
\end{figure}

While these calculations are strictly valid only for small local stresses, one may hope that Eq.~\eqref{eq:Pa=f(Pl)} remains qualitatively correct for strong active units with $\mathcal P_l\approx \mathcal S_l\gtrsim\kappa$. We test this expectation through finite element simulations [SI] of a fully (\emph{i.e.}, not weakly) nonlinear model with an elastic energy density
\begin{equation}
	E=\frac{\kappa}{2}\frac{\left(J-1\right)^2}{1+a(J-1)} + \frac{\mu}{2}\frac{I/J-2}{1+b(J-1)},
\end{equation}
where $J=\det(\mathbf1+\bm\eta)$, $I=\Tr(\mathbf1+2\bm\varepsilon)$ and the constants $a,b$ are defined through $\kappa_1=1/2-3a$, $\mu_1=-3/2-b$. The case $a=b=0$ corresponds to a compressible neo-Hookean model for rubber elasticity. We illustrate a bucklable and an anti-bucklable material in Fig.~\ref{FES_Shokef} by choosing two media with $\kappa_1=\mu_1=1$ and $\kappa_1=\mu_1=-4$ (equidistant from $-3/2$, as denoted by symbols in Fig.~\ref{K1M1_full}(a)). As expected, the former induces contraction while the latter causes expansion. The quantitative predictions of Eq.~\eqref{eq:Pa=f(Pl)} moreover remain largely valid up to local stress values comparable with the bulk modulus of the network, which implies deformations of the medium of order one. These conclusions also hold in other parameter regimes and for a model specifically designed to mimic the shear-stiffening behavior of fiber networks (Fig.~S4)~\cite{gardel_elastic_2004}. In addition, simulations of isotropic active units with large local stress values suggest that rectification effects also manifest in that case (Fig.~S5)~\cite{ronceray_fiber_2016}.

Our intuition of the mechanics of nonlinear materials is largely based on deforming their outer boundary. We thus expect a uniformly compressed material to respond with an expansile stress, while applying shear will elicit an opposing shear stress. In this study, we show that if the forces are exerted from the \emph{inside} of the material, these expectations can be upset. In the most extreme cases, an embedded active unit that expands (contracts) in all directions can elicit contractile (expansile) stresses in all directions. The system thus ``forgets'' the shape of the active units, and its large-scale behavior is controlled by the characteristics of the elastic material instead. Expansion- and shear-stiffening (softening) materials thus always rectify towards contraction (expansion). This rectification tends to be stronger in more compressible materials and in larger systems. These behaviors arise in a continuum model with or without constitutive nonlinearities, and are thus generic in elastic media beyond previously studied discrete fiber networks. 

While most of our calculations are conducted in a circular 2D system with a single active unit, they are likely to remain valid in more complex settings provided the elastic medium is homogeneous. Indeed, Refs.~\cite{ronceray_connecting_2015,ronceray_fiber_2016} show that if an active unit is far enough away from the boundary of the medium and from other active units, its contribution to the total active stress is independent of the characteristics of either. This remains true as long as the distance between active units is larger than the distance over which each of them induces significant nonlinear deformations. In our small-strain formalism (which also describes intermediate strains well), this distance is of the order of $r^*\sim10\,r_\text{in}$ [SI].

In the strongly nonlinear regime, rectification in fiber networks is strikingly similar to the results of our weakly nonlinear formalism~\cite{ronceray_fiber_2016}, which may explain why actomyosin networks are always contractile despite the presence of mixed force dipoles~\cite{Hatano:1994, Lenz_NJoP12}. Its application to discrete granular media and other amorphous solids remains to be investigated. Experiments do however suggest that the elastic response of a foam to a shear transformation zone becomes more isotropic in the vicinity of the jamming transition~\cite{Desmond15}, where nonlinear effects are expected to play a large role. We speculate that such effects could be explained by the type of rectification described here. They could then significantly affect the characteristics of the yielding transition in nearly-jammed systems~\cite{Nicolas18, Merabia16}.

\acknowledgments{We thank Pierre Ronceray and Mehdi Bouzid for many discussions and suggestions, and Lev Truskinovsky for comments on the manuscript. ML was supported by Marie Curie Integration Grant PCIG12-GA-2012-334053, “Investissements d’Avenir” LabEx PALM (ANR-10-LABX-0039-PALM), ANR grants ANR-15-CE13-0004-03 and ANR-21-CE11-0004-02, as well as ERC Starting Grant 677532. ML’s group belongs to the CNRS consortium AQV.}

\end{document}


\title{\Large{Supporting information for ``Generic stress rectification in nonlinear elastic media''}}
\maketitle

\section{Constitutively linear media rectify towards contraction}\label{1}
In this section, we consider only geometrical nonlinearities and set out to prove the inequality in Eq.~(3). Firstly, the elastic energy density $E$ is written with the Green-Lagrange strain tensor $\bm\varepsilon=(\bm\eta+\bm\eta^\mathsf T+\bm\eta^\mathsf T\bm\eta)/2$, which depends quadratically on the displacement gradient $\eta_{ij}=\partial u_i/\partial x_j$. We further define the symmetric part of the displacement gradient $U_{ij}=(\eta_{ij}+\eta_{ji})/2$ which corresponds to the linearized strain. The stress tensor which naturally derives from $E$ describes the surface force measured in the initial space with respect to the initial area: $\text d\mathbf F^0/\text d\mathbf a=\partial E/\partial\bm\varepsilon$, it is known as the second Piola-Kirchhoff stress. Then in order to find the Cauchy stress measured fully in the target space: $\bm\sigma=\text d\mathbf F/\text d\mathbf A$, we need to transform the surface force and the area as 
\begin{equation}
	\text d\mathbf F=(\mathbf1+\bm\eta)\,\text d\mathbf F^0\qq{ and }\text d\mathbf a=\frac{\mathbf1+\bm\eta^\mathsf T}{\det(\mathbf1+\bm\eta)}\text d\mathbf A,
\end{equation}
which ultimately gives the formula for the Cauchy stress in the main text~\cite{Wriggers08}. Then, given the quadratic energy density $E=\kappa\varepsilon_{ii}^2/2+\mu(\varepsilon_{ij}^2-\epsilon_{ii}^2/d)$, the stress-strain relation displays a linear stress term $\bm\sigma^L$ proportional to $\bf U$, and a term which includes the geometrical nonlinearities $\bm\sigma^G$:
\begin{subequations}\label{eq:sig_decomp}
\begin{align}\label{eq:sig_LGC}
	\bm\sigma^L&=\big(\kappa-2\mu/d\big)U_{ii}\,\mathbf1+2\mu\,\mathbf{U},\\[5pt]
	\bm\sigma^G&=\big(\kappa-2\mu/d\big)\big(\eta_{ij}^2/2-U_{ii}^2\big)\mathbf1+\mu\big(4\mathbf U^2+\bm\eta\bm\eta^\mathsf T\big)+2\big(\kappa-2\mu/d-\mu\big) U_{ii}\mathbf U+\order{\eta^3}.
\end{align}
\end{subequations}

Secondly, given the force balance equation $f_i=-\partial\sigma_{ij}/\partial X_j$, the difference between the local and active coarse-grained stresses can be integrated by part to read
\begin{equation}\label{eq:mst}
	\bar{\bm\sigma}^a-\bar{\bm\sigma}^l=\frac{1}{V}\int_{\Omega}\bm\sigma\,\text{d}V=\frac{1}{V}\int_{\Omega}\bm\sigma\det(\mathbf1+\bm\eta)\,\text{d}v,
\end{equation}
where $\text{d}v$ is the volume element in the initial space. Equation~\eqref{eq:mst} is known as the mean stress theorem~\cite{ronceray_connecting_2015, carlsson_contractile_2006}. Due to this relation, writing $\mathbf S=\bm\sigma\det(\mathbf1+\bm\eta)$ and decomposing $\mathbf S=\mathbf S^L+\mathbf S^G$ as we did $\bm\sigma$ in Eq.~\eqref{eq:sig_decomp}, the pressure difference reads $P_a-P_l=-\int S_{ii}/(Vd)$. Due to our fixed boundary condition, the integral of the trace of the linear term $\mathbf S^L=\bm\sigma^L$ vanishes and the trace of the nonlinear term is expressed in a closed form as
\begin{equation}\label{eq:tr_sig_GC}
	S_{ii}^G=\frac{\kappa}{2}\left(d\eta_{ij}^2+4\varepsilon_{ii}^2\right)+4\mu\big(\varepsilon_{ij}^2-\varepsilon_{ii}^2/d\big).
\end{equation}
Here, $\kappa$ and $\mu$ are both positive for mechanical stability. Thus, for $d\ge2$, the geometrical term $S^G_{ii}$ always gives a positive (contractile) contribution to the active stress. Indeed, using the eigenvalues $\lambda_i\in 	\mathbb{R}$ of the symmetric matrix $\bm\varepsilon$, we can rewrite $\varepsilon_{ij}^2-\varepsilon_{ii}^2/d=\sum_{i<j}(\lambda_i-\lambda_j)^2/d$, which is always non-negative. As a result $S^G_{ii}$ is a sum of squares that is also non-negative, implying the inequality of Eq.~(3): $P_a-P_l\leqslant 0$.

Finally, we present an alternative derivation of this relation that highlights its frame indifference. The integrand of Eq.~(3) can be rewritten using the deformation gradient $\bm\Lambda=\mathbf1+\bm\eta$ and the right Cauchy-Green deformation tensor $\mathbf C=\bm\Lambda^\mathsf T\bm\Lambda=\mathbf 1+2\bm\varepsilon$. Indeed, since
\begin{equation}\label{eq:frame}
	\mathbf S=\bm\Lambda\frac{\partial E}{\partial\bm\varepsilon}\bm\Lambda^\mathsf T, \qq{where} \frac{\partial E}{\partial\bm\varepsilon}=\frac{\kappa}{2}(C_{ii}-d)\mathbf1+\mu\big(\mathbf C-C_{ii}\mathbf1/d\big),
\end{equation}
we can write
\begin{equation}\label{eq:frame1}
\begin{aligned}
	S_{ii}=\frac{\kappa}{2}\left[(C_{ii}-d)^2+dC_{ii}\right]+\mu\big[C_{ij}^2-C_{ii}^2/d\big].
\end{aligned}
\end{equation}
We see that the right-hand side Eq.~\eqref{eq:frame1} is a sum of squares plus a term $\propto C_{ii}=d+4\eta_{ii}+4\eta_{ij}^2$. Since $\eta_{ii}$ is integrated to zero due to the fixed boundary condition, the term in Eq.~\eqref{eq:frame1} also gives a contractile contribution to $P_a-P_l$.

\section{Elastic moduli in granular media near jamming}

\begin{figure}[!t]
    \centering
	   \begin{tikzpicture}[scale=1.]
    	\node[anchor=north] at (0,0)
   		{ \includegraphics[width=.2\textwidth]{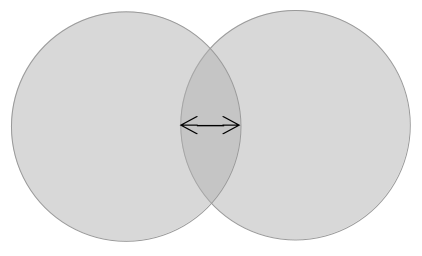}};
   		\draw (-.22,-.84) node[right]{\scalebox{1.2}{$\delta$}};
	\end{tikzpicture}
    \caption{Overlap between two interacting beads in a granular simulation.}
   \label{overlap}
\end{figure} 

Here, we derive typical values of $\kappa_1$ and $\mu_1$ for granular media near the jamming transition.
Let us consider a large equilibrium packing of bidisperse frictionless spherical grains in a 2D box with volume fraction $\phi$. The grains interact through a harmonic potential $\mathcal V\sim k\,\delta^2$, where $k$ is a spring constant, and $\delta$ the overlap divided by the sum of the two bead diameters, see Fig.~\ref{overlap}. For volume fractions slightly above the jamming transition $\phi_c\approx0.84$, granular media display a strongly nonlinear elastic behavior. Indeed, multiple simulations and experiments~\cite{O'Hern03,vanHecke09} have shown that while the bulk modulus goes to a finite limit in $\phi_c^+$ and can thus be approximated by a constant, the shear modulus scales with $\phi-\phi_c$ and vanishes at the transition. This can be expressed as
\begin{equation}\label{eq:K,G_gran}
	K/k\sim K_0\qq{and} G/k\sim G_0(\phi-\phi_c)^p,\qq{for} 0<\phi-\phi_c\ll1,
\end{equation}
where $K_0, G_0\approx0.2$ and $p\approx0.5$. Based on this model, we impose an isotropic compression characterized by a displacement gradient tensor $\eta_{ij}=-\eta_0\delta_{ij}/d$ on our granular material initially at $\phi_c$, where $0<\eta_0\ll1$. This results in a new volume fraction $\phi_0$ such that
\begin{equation}
	\phi_0-\phi_c=\phi_0\eta_0=\phi_c\frac{\eta_0}{1-\eta_0}>0.
\end{equation}
We then compute the elastic moduli $\kappa$ and $\mu$, and their nonlinear corrections $\kappa_1$ and $\mu_1$ around this value of $\phi_0$.

Let the bulk strain $\eta_{ii}=-\eta_0+\delta\eta$, where $\abs{\delta\eta}\ll\eta_0$, corresponding to a volume fraction $\phi\sim\phi_0-\phi_c\delta\eta$. Then similarly to Eq.~(6), the moduli are expressed as $K\sim \kappa(1+\kappa_1\delta\eta)$ and $G\sim \mu(1+\mu_1\delta\eta)$, where the parameters are derived from Eq.~\eqref{eq:K,G_gran}:
\begin{equation}
\begin{aligned}
	\kappa/k &=K_0,\hspace{2.5cm}  \kappa_1=0, \\
\mu/k &= G_0\left(\frac{\phi_c\eta_0}{1-\eta_0}\right)^p,\qquad \mu_1=-\frac{p}{\eta_0(1-\eta_0)}.
\end{aligned}
\end{equation}
Therefore, while $\kappa_1$ vanishes, $\mu_1$ diverges at the transition and scales as $(\phi_0-\phi_c)^{-1}$. And close to the transition, around \emph{e.g.} $\phi_0-\phi_c= 0.001,\,0.01$ or 0.1, we find respectively $\mu_1\approx-400,\,-40$ or $-4$ as in the main text.

\section{Coarse-grained stresses in the circular geometry}
In this section, we present the analytical calculations leading to the expressions of the coarse-grained stresses $\bar{\bm\sigma}$ in Eq.~(9) and of the rectification coefficient $\alpha$ of Eq.~(10) of the main text. We first rewrite the coarse-grained stresses in the initial space where the calculations will be easier to handle in Sec.~\ref{3a}. Then in Sec.~\ref{3b} we present the Ansatz for the displacement field that allows us to solve the force balance condition. In Sec.~\ref{3c}, we show the expressions of the coarse-grained pressures and shear stresses with the stress and displacement fields. We finally display the detailed expressions of the coarse-grained stresses $\bar{\bm\sigma}$ in the circular geometry as well as the expression of $\alpha$ in Sec.~\ref{3d}.

	\subsection{Rewriting the coarse-grained stresses in the initial space}\label{3a}
In order to calculate the coarse-grained stresses, we need to express them in the initial configuration, where the force density $\bm\phi=\mathbf f/\det(\mathbf1+\bm\eta)$ is related to the first Piola-Kirchhoff stress tensor $\bm\tau=(\mathbf1+\bm\eta)\frac{\partial E}{\partial\bm\varepsilon}$. The force balance equation thus reads $\phi_i=-\partial_j\tau_{ij}$ and the coarse-grained stresses can be rewritten as
\begin{equation}\label{eq:barsig_tau}
	\bar{\sigma}^a_{ij}=\frac{1}{V}\oint_{\partial\Omega}\tau_{ik}X_j \,\text{d}a_k \qq{and} \bar{\sigma}^l_{ij}=-\frac{1}{V}\int_\Omega \phi_iX_j\,\text{d}v,
\end{equation}
where $\text{d}\mathbf{a}$ is the outward-directed area element in the initial space. As before in Sec.~\ref{1}, we distinguish the linear term $\bm\tau^L$ (which is the same in the initial and target spaces $\bm\tau^L=\bm\sigma^L$), from the nonlinear term $\bm\tau^\text{NL}=\bm\tau^G+\bm\tau^C$. This last equality distinguishes the geometrical and constitutive nonlinearities. Given the non-harmonic energy density of Eq.~(4), up to second order, the terms read
\begin{subequations}\label{eq:tau_LGC}
\begin{align}
	\bm\tau^L&=\big(\kappa-2\mu/d\big)U_{ii}\,\mathbf1+2\mu\,\mathbf{U},\\[5pt]
	\bm\tau^G&=\big(\kappa-2\mu/d\big)\eta_{ij}^2\,\mathbf1/2+\mu\,\bm\eta^\mathsf T\bm\eta+\big(\kappa-2\mu/d\big)U_{ii}\,\bm\eta+2\mu\,\bm\eta\mathbf U+\order{\eta^3}.\\[5pt]
	\bm\tau^C&=3\big(\kappa'-2\mu'/d\big)U_{ii}^2\,\mathbf1/2+\mu'U_{ij}^2\,\mathbf1+2\mu'U_{ii}\,\mathbf U+\order{\eta^3}.
\end{align}
\end{subequations}

	\subsection{Ansatz for the displacement field}\label{3b}
In the initial space, the material is subjected to zero body force except at $r_\text{in}$, where the stress is discontinuous. The fixed boundary at $r_\text{out}$ and the imposed displacement at $r_\text{in}$ additionally impose boundary conditions on the stress and displacement fields, resulting in the following system of equations:
\begin{equation}\label{eq:system}
\left\{\begin{aligned}
	\div\bm\tau^\mathsf T&=\mathbf0, \quad\text{for } r\in[0,r_\text{in})\cup (r_\text{in},r_\text{out})\\
	\mathbf u&=\mathbf0, \quad\text{at } r=0 \text{ and } r=r_\text{out}\\
	\mathbf u&=r_\text{in}\left[e_0+e_2\cos(2\theta)\right]\hat{\mathbf r}, \quad\text{at } r=r_\text{in}\\
\end{aligned}\right..
\end{equation}
We solve this system perturbatively by expanding $\bf u,\bm{\eta,\tau}$ in the small scalar quantity
\begin{equation}
	\eta\sim \abs{e_0}+\abs{e_2}.
\end{equation}
The displacement gradient is hence written $\bm\eta = \bm\eta^L+\bm\eta^\text{NL} +\order{\eta^3}$ where the $L$ and $\text{NL}$ superscript refer to linear and quadratic (nonlinear) terms in $\eta$. This allows us to write the stress tensor as
\begin{equation}
\begin{aligned}\label{eq:expansion}
	\bm\tau &= \bm\tau^L\big(\bm\eta^L\big)+\bm\tau^L\big(\bm\eta^\text{NL}\big)+\bm\tau^\text{NL}\big(\bm\eta^L\big) +\order{\eta^3},
\end{aligned}
\end{equation}
where $\bm\tau^L\big(\bm\eta^L\big)$ is of order 1, while the next two are of order 2. The linear displacement field $\mathbf u^L$ is the solution of $\partial_i\tau_{ji}^L(\bm\eta^L)=0$. This is solved by decomposing $\mathbf u^L$ in the following Fourier modes due to the form of the imposed displacement:
\begin{subequations}\label{eq:u^L}
\begin{align}
	u_r^L/r_\text{in}&=e_0\zeta_0(r)+e_2\zeta_2(r)\cos 2\theta,\\[5pt]
u_\theta^L/r_\text{in}&=e_2\omega_2(r)\sin 2\theta.
\end{align}
\end{subequations}
Here $\zeta_0(r)$, $\zeta_2(r)$, $\omega_2(r)$ are sums of $r^k$, with $k\in\{-3,-1,1,3\}$ and coefficients depending on the boundary conditions. Then, at the first nonlinear order, $\mathbf u^\text{NL}$ is the solution of the linear equation $\partial_i\tau_{ji}^L(\bm\eta^\text{NL})=-\partial_i\tau_{ji}^\text{NL}(\bm\eta^L)$ which is solved by expanding $\mathbf u^\text{NL}$ as
\begin{subequations}\label{eq:u^NL}
\begin{align}
	u_r^\text{NL}/r_\text{in}&=e_2^2\xi_0(r)+e_0e_2\xi_2(r)\cos2\theta+e_2^2\xi_4(r)\cos4\theta,\\[5pt]
	 u_\theta^\text{NL}/r_\text{in}&=e_0e_2\pi_2(r)\sin2\theta+e_2^2\pi_4(r)\sin4\theta,
\end{align}
\end{subequations}
where the $\xi_i(r)$, $\pi_i(r)$ are again sums of $r^k$ with $k$ odd between $-7$ and $+5$. As a consequence, we obtain in Sec.~\ref{3d} the strain and stress fields up to second order in $\eta$.

	\subsection{Calculations of the coarse-grained stresses}\label{3c}
We compute the coarse-grained active stress $\bar{\bm{\sigma}}^a$ and local stress $\bar{\bm{\sigma}}^l$ in the circular geometry by integrating the stresses in the material as in Eq.~\eqref{eq:barsig_tau}. The coarse-grained stresses are expressed in Cartesian coordinates, while the stress field is more easily expressed in polar coordinates. In the following (and in this subsection only), we denote Cartesian indices $x,y$ with Greek letters ($\mu,\nu$) and polar indices $r,\theta$ with Latin letters ($i,j,k$). The change-of-basis matrix between these two systems reads $\mathbf R=\left(\begin{smallmatrix}\cos\theta&-\sin\theta\\\sin\theta&\cos\theta\end{smallmatrix}\right).$ In the circular geometry of Fig.~1 where the active unit produces a discontinuity in the stress at $r_\text{in}$ in the initial configuration, the coarse-grained stresses are expressed as follows:
\begin{subequations}
\begin{align}
	\bar\sigma^a_{\mu\nu}&=\frac{1}{\pi r_\text{out}^2}\oint_{\partial\Omega} R_{\mu i}\,\tau_{ik}\, R_{\nu j}\,X_j \,\text{d}a_k = \frac{1}{\pi}\int_0^{2\pi}\text{d}\theta\,R_{\mu i}\,\tau_{ir}(r_\text{out},\theta)R_{\nu r},\\
	\bar\sigma^l_{\mu\nu}&=\frac{1}{\pi r_\text{out}^2}\int_{\Omega}R_{\mu i}\,\partial_k\tau_{ik}\,R_{\nu j}\,X_j\,\text{d}V = \frac{1}{\pi}\frac{r_\text{in}^2}{r_\text{out}^2}\int_0^{2\pi}\text{d}\theta\,R_{\mu i}\Big[\tau_{ir}(r_\text{in}^+,\theta)-\tau_{ir}(r_\text{in}^-,\theta)\Big]R_{\nu r}\left[1+\frac{u_r(r_\text{in},\theta)}{r_\text{in}}\right],
\end{align}
\end{subequations}
where $\partial_k\tau_{ik}$ denotes the stress divergence expressed in polar coordinates. Then, introducing $\beta=(r_\text{out}/r_\text{in})^2$, the active pressure and shear stress rescaled so as to compensate for dilution read
\begin{subequations}\label{eq:P_a}
\begin{align}
	\beta(\bar\sigma^a_{xx}+\bar\sigma^a_{yy})=-2\mathcal P_a&=\frac{\beta}{\pi}\int\tau_{rr}(r_\text{out},\theta),\\
	\beta(\bar\sigma^a_{xx}-\bar\sigma^a_{yy})=-2\mathcal S_a&=\frac{\beta}{\pi}\int\Big[\tau_{rr}(r_\text{out},\theta)\cos{2\theta}-\tau_{\theta r}(r_\text{out},\theta)\sin{2\theta}\Big],
\end{align}
\end{subequations}
with similar expressions for the local pressure and shear stress:
\begin{subequations}\label{eq:P_l}
\begin{align}
	-2\mathcal P_l&=\frac{1}{\pi}\int\Big[\tau_{rr}(r_\text{in}^+,\theta)-\tau_{rr}(r_\text{in}^-,\theta)\Big]\left[1+\frac{u_r(r_\text{in},\theta)}{r_\text{in}}\right],\\
	-2\mathcal S_l&=\frac{1}{\pi}\int\bigg\{\Big[\tau_{rr}(r_\text{in}^+,\theta)-\tau_{rr}(r_\text{in}^-,\theta)\Big]\cos{2\theta}-\Big[\tau_{\theta r}(r_\text{in}^+,\theta)-\tau_{\theta r}(r_\text{in}^-,\theta)\Big]\sin{2\theta}\bigg\}\left[1+\frac{u_r(r_\text{in},\theta)}{r_\text{in}}\right].
\end{align}
\end{subequations}

	\subsection{Full expressions of the coarse-grained stresses}\label{3d}
In Eqs.~(9-10), we consider small displacements with two independent parameters $e_0,e_2\ll1$. In the weakly nonlinear formalism, the
simplest possible rectification requires that the term in $e_0$ be similar (and of opposite sign) to the term in $e_2^2$. In this regime, given $\epsilon\ll1$, the displacement parameters read $e_0=\epsilon\, \tilde e_0$ and $e_2=\sqrt\epsilon\,\tilde e_2$. As a result $\mathcal P_x=\epsilon \tilde{\mathcal P}_x$ and $\mathcal S_x=\sqrt{\epsilon} \tilde{\mathcal S}_x$ for $x\in\{l,a\}$, where the tildes denote quantities of order one. Then to lowest order in $\epsilon$, Eq.~(9) can be rewritten as
\begin{subequations}
\begin{align}
	\tilde{\mathcal P}_x&= A_x\tilde e_0+ B_x\tilde e_2^2 +\order{\epsilon},\\
	\tilde{\mathcal S}_x&= C_x\tilde e_2 +\order{\epsilon},
\end{align}
\end{subequations}
implying that the ``$\sim$'' symbols of Eq.~(10) denote equalities to lowest order in $\epsilon$. The rectification behavior thus depends on the ratio $\tilde{\mathcal P}_l/(\alpha\tilde{\mathcal S}_l^2)$, \emph{i.e.} on $\tilde e_0/\tilde e_2^2$. We further display the complete expressions of the coefficients in Eq.~(9) obtained after finding the strain field through force balance (see Sec.~\ref{3b}) and integrating the resulting stress via Eq.~\eqref{eq:tau_LGC}, as in \eqref{eq:P_a} and \eqref{eq:P_l}. In order to make sense of the cumbersome expressions of the $A_x, B_x$ and $C_x$, we introduce several quantities: $X=(3-\nu)^2(1+\beta^2)+2(3-6\nu-\nu^2)\beta$,
\begin{align*}
	a_0&= 1215-1863\nu+756\nu^2-126\nu^3+87\nu^4-43\nu^5+6\nu^6\\
&\quad +(81-2457\nu+2412\nu^2-594\nu^3+265\nu^4+155\nu^5-22\nu^6)\beta\\
&\quad +(1782-3798\nu+6216\nu^2-3948\nu^3+326\nu^4-286\nu^5+28\nu^6)\beta^2\\
&\quad -(918+1170\nu-2616\nu^2+1044\nu^3+54\nu^4-262\nu^5+12\nu^6)\beta^3\\
&\quad +(459+45\nu+132\nu^2-918\nu^3+515\nu^4-71\nu^5-2\nu^6)\beta^4\\
&\quad -(891-1755\nu+1188\nu^2-294\nu^3-13\nu^4+17\nu^5-2\nu^6)\beta^5,\\[5pt]
	a_1&= 1863-756\nu-54\nu^2+12\nu^3+7\nu^4+(1377-1260\nu+102\nu^2-108\nu^3-31\nu^4)\beta\\
&\quad +(1782-1368\nu+804\nu^2+168\nu^3+54\nu^4)\beta^2+(162-504\nu-132\nu^2-24\nu^3-46\nu^4)\beta^3\\
&\quad +(243+108\nu+18\nu^2-84\nu^3+19\nu^4)\beta^4-(243-324\nu+162\nu^2-36\nu^3+3\nu^4)\beta^5,
\end{align*}
\begin{align*}
	a_2&= 2511-4104\nu+4023\nu^2-1416\nu^3+173\nu^4-48\nu^5+13\nu^6\\
&\quad +(1377-8352\nu+6369\nu^2-2568\nu^3-77\nu^4+104\nu^5-53\nu^6)\beta\\
&\quad+(6966-16128\nu+17430\nu^2-6480\nu^3+1890\nu^4-176\nu^5+82\nu^6)\beta^2\\
&\quad+(2754-9216\nu+6162\nu^2-1392\nu^3-794\nu^4+240\nu^5-58\nu^6)\beta^3\\
&\quad+(2187-3672\nu+3843\nu^2-1896\nu^3+545\nu^4-128\nu^5+17\nu^6)\beta^4\\
&\quad -(243-189\nu^2+72\nu^3+9\nu^4-8\nu^5 +\nu^6)\beta^5,
\end{align*}
and
\begin{align*}
	b_0&= 81-270\nu+234\nu^2-36\nu^3+13\nu^4-6\nu^5+(135-378\nu+366\nu^2-196\nu^3-29\nu^4+22\nu^5)\beta \\
&\quad+(162-708\nu+1116\nu^2-464\nu^3+82\nu^4-28\nu^5)\beta^2+(126-564\nu+540\nu^2-168\nu^3-106\nu^4+12\nu^5)\beta^3 \\
&\quad+(45-222\nu+426\nu^2-188\nu^3+17\nu^4+2\nu^5)\beta^4+(27-162\nu+198\nu^2-100\nu^3+23\nu^4-2\nu^5)\beta^5,\\[5pt]
	b_1&= 27-18\nu+3\nu^2+(189+18\nu-11\nu^2)\beta+(174+12\nu+14\nu^2)\beta^2\\
&\quad +(138-12\nu-6\nu^2)\beta^3+(39+6\nu-\nu^2)\beta^4+(9-6\nu +\nu^2)\beta^5,\\[5pt]
	b_2&= 243-270\nu+84\nu^2-34\nu^3+9\nu^4+(189-522\nu+404\nu^2+58\nu^3-33\nu^4)\beta\\
&\quad +(510-1356\nu+568\nu^2-84\nu^3+42\nu^4)\beta^2+(474-612\nu+312\nu^2+100\nu^3-18\nu^4)\beta^3\\
&\quad +(159-486\nu+244\nu^2-10\nu^3-3\nu^4)\beta^4+(153-210\nu+116\nu^2-30\nu^3+3\nu^4)\beta^5.
\end{align*}
In the end, we find
\begin{subequations}\label{eq:A_l}
\begin{align}
	A_l=A_a&= \frac{4\kappa\beta}{(1+\nu)(\beta-1)},\\[5pt]
	B_l&= -\kappa\beta \frac{(3-\nu)a_0+(1-\nu)^2(1+\nu)a_1\,\kappa_1 +(1-\nu)a_2\,\mu_1}{(3-\nu)^2(1+\nu)(\beta-1)^2X^2},\\[5pt]
	B_a&= -\kappa\beta \frac{b_0+(1-\nu)^2(1+\nu)b_1\,\kappa_1 +(1-\nu)b_2\,\mu_1}{(1+\nu)(\beta-1)^2X^2},\\[5pt]
	C_l=C_a&= 4\mu\beta \frac{2 (3 +\nu)+(3-\nu)(\beta+\beta^2)}{(\beta-1)X}.
\end{align}
\end{subequations}
As expected from the linear elasticity analysis of Sec.~\ref{1}, the coefficients in front of the linear $e_0$ and $e_2$ terms are identical for the local and boundary stresses, but discrepancies appear in the $e_2^2$ terms. This leads us to define $\alpha=(B_a-B_l)/C_l^2$.

\section{Behavior of \texorpdfstring{$\alpha$}{\textit{alpha}} and rectification saturation radius \texorpdfstring{$r^*$}{\textit{r*}}}
\begin{figure}[!t]
    \centering
	\begin{tikzpicture}[scale=1.325]
    	\node[anchor=north west,xshift=+1.55mm,yshift=-1.25mm] at (0,0)
   		{\includegraphics[width=.6\linewidth]{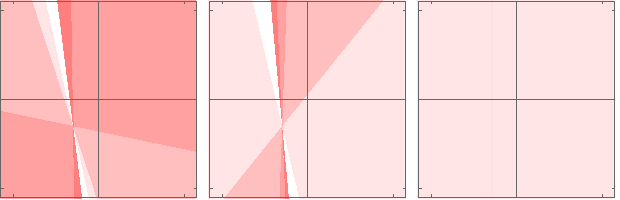}};
   		\draw(2.6,-2.3) node[anchor=east,rotate=-12.]{\scalebox{.9}{$\sqrt\beta_{10}>5$}};
   		\draw(2.65,-1.98) node[anchor=east,rotate=-12.]{\scalebox{.9}{$\sqrt\beta_{10}<5$}};
    	\draw [->,thick] (0,0)--(8.6,0) node (xaxis) [right] {$\nu$};
	    \draw(1.49,1pt)--(1.49,-2pt) node[anchor=south,yshift=+1mm]{0};					\draw(4.22,1pt)--(4.22,-2pt) node[anchor=south,yshift=+1mm]{0.5}; 					\draw(6.96,1pt)--(6.96,-2pt) node[anchor=south,yshift=+1mm]{1};	
    	\draw[gray] (8.23,-.31) node[right]{5};
    	\draw[gray] (8.23,-1.49) node[right]{0};
    	\draw[gray] (8.23,-2.63) node[right]{-5};
    	\draw[gray] (8.53,-1.49) node[right]{$\kappa_1$};
    	\foreach \x in {0,1,2}
    		{\draw[gray] (\x*2.75+.35,-2.76) node[below]{-5};				 			\draw[gray] (\x*2.75+1.49,-2.76) node[below]{0};
    	 	\draw[gray] (\x*2.75+2.62,-2.76) node[below]{5};}
   		\draw[gray] (2.75+1.49,-3.07) node[below]{$\mu_1$};
	\end{tikzpicture}
	\begin{tikzpicture}[scale=1.325]
    	\node[anchor=north west,xshift=+1.55mm,yshift=-1.25mm] at (0,0)
   		{\includegraphics[width=.019\linewidth]{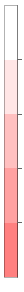}};
   		\draw(.4,-.3) node[right]{$\sqrt\beta_{10}$};	
   		\draw(.4,-.8) node[right]{9};
   		\draw(.4,-1.4) node[right]{6};
   		\draw(.4,-1.98) node[right]{5};
   		\draw(.4,-2.6) node[right]{4};
   		\draw(.4,-3.15)--(.45,-3.15) node[anchor=west,xshift=-.5mm]{1};
   		\draw[white](.4,-3.5)node[right]{1};
	\end{tikzpicture}	
    \caption{The rectification saturation radius $r^*$ is generally of the order of $r_\text{in}$. Contour plot of $\sqrt\beta_{10}=r^*/r_\text{in}$ when $\kappa_1$ and $\mu_1$ are varied for several values of Poisson's ratio $\nu$. Except in the small white regions, $\sqrt\beta_{10}<9$. In these white regions, $|\alpha|$ tends to take negligible values.}
\label{beta10}
\end{figure}

\begin{figure}[!t]
    \centering
	\begin{tikzpicture}[scale=1.325]
    	\node[anchor=north west,xshift=+1.4mm,yshift=-.75mm] at (0,0)
   		{\includegraphics[width=.7\linewidth]{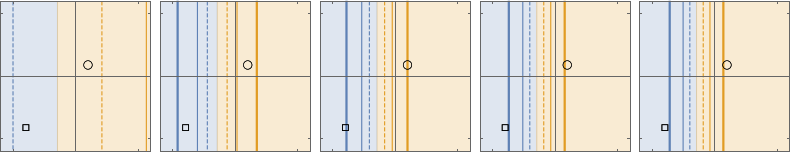}};
    	\draw [->,thick] (0,0)--(9.99,0) node (xaxis) [right] {$\dfrac{r_\text{out}}{r_\text{in}}$};
	    \draw(1.085,1pt)--(1.085,-2pt) node[anchor=south,yshift=+1mm]{2};				\draw(3.00,1pt)--(3.00,-2pt) node[anchor=south,yshift=+1mm]{4}; 				\draw(4.925,1pt)--(4.925,-2pt) node[anchor=south,yshift=+1mm]{6};				\draw(6.84,1pt)--(6.84,-2pt) node[anchor=south,yshift=+1mm]{8}; 				\draw(8.75,1pt)--(8.75,-2pt) node[anchor=south,yshift=+1mm]{10};
    	\draw[gray] (9.62,-.31) node[right]{5};
    	\draw[gray] (9.62,-1.07) node[right]{0};
    	\draw[gray] (9.62,-1.83) node[right]{-5};
    	\draw[gray] (9.92,-1.07) node[right]{$\kappa_1$};
    	\foreach \x in {0,1,2,3,4}
    		{\draw[gray] (\x*1.915+.15,-2.09) node[right]{-5};				 			\draw[gray] (\x*1.915+.942,-2.09) node[right]{0};
    	 	\draw[gray] (\x*1.915+1.69,-2.09) node[right]{5};}
   		\draw[gray] (2*1.915+.88,-2.41) node[right]{$\mu_1$};
	\end{tikzpicture}
    \caption{Rectification happens mostly near the active unit. Additional plots to Fig.~2(a) showing the stabilization of the graphs as the system size increases for a Poisson's ratio $\nu=1$. The blue and yellow lines at constant $\abs{\alpha}\mu$ stay quite still between $r_\text{out}/r_\text{in}=8$ and 10. The behavior is similar for $\nu=0$.}
\label{K1M1_SI}
\end{figure}

To help better understand the lengthy expression of the rectification coefficient $\alpha=(B_a-B_l)/C_l^2$ in subsection III.D, we hereby discuss its dependence on the system size $r_\text{out}$. As is apparent from Fig.~2(a), $\alpha$ increases with increasing $r_\text{out}$ for relatively small systems, then saturates as the size of the system goes to infinity. Indeed, away from the high-stress region close to the active unit, the stress decrease causes the nonlinearities to become negligible in front of the linear terms. Therefore, we examine the radius $r^*$ at which stress propagation switches from nonlinear to linear. To this end we define the system size parameter $\beta=(r_\text{out}/r_\text{in})^2$. In the limit $\beta \to\infty$, Eq.~\eqref{eq:A_l} leads to $\alpha(\beta)=\alpha_\infty+\alpha_1/\beta +\mathcal{O}(\beta^{-2})$, where
\begin{equation}
\begin{aligned}
	\mu\alpha_\infty&= -\frac{\big(\kappa_1+\tfrac{3}{2}\big)(1-\nu ^2)+\big(\mu_1+\tfrac{3}{2}\big)(5-2 \nu+\nu ^2)}{4},\\
\mu\alpha_1&=\frac{\big(\kappa_1+\tfrac{3}{2}\big)(15-6\nu-8\nu^2+6\nu^3-7 \nu ^4)+ \big(\mu_1+\tfrac{3}{2}\big)(111-36\nu+50\nu^2-20\nu^3+7 \nu ^4)}{4(3-\nu)^2}.
\end{aligned}
\end{equation}
We introduce the value of the parameter $\beta$ such that $\alpha$ is within $10\%$ of its large-size limit through $\abs{\frac{\alpha(\beta_{10})-\alpha_\infty}{\alpha_\infty}}=0.1$. 
The square-root of $\beta_{10}$, corresponding to the ratio of the radii, lies between 4 and 9, except for insignificant values of $\alpha$, see Fig.~\ref{beta10}. Fig.~\ref{K1M1_SI} also illustrates this behavior by showing the stabilization of the lines at constant $\alpha$ as the system size increase. Therefore, defining the rectification saturation radius $r^*$ such that $\beta_{10}=(r^*/r_\text{in})^2$, increasing $r_\text{out}$ past $r^*\sim10 r_\text{in}$ has little influence on the value of $\mathcal P_a-\mathcal P_l$, \emph{i.e.} on the rectification effect. This indicates that the propagation is nonlinear only up to $r^*$. In the study of stress propagation from multiple active units, one thus needs to compare this $r^*$ to the typical spacing between two active units. 

\section{The rectification diagram}
Let us give further explanation to the shadings of the rectification diagram of Fig.~2(b), corresponding to different rectification regimes. In the circular geometry of Fig.~1, provided that $\alpha$ and $\mathcal P_l$ have different signs, a change of sign of $P_a$ due to rectification can appear for all values of the local pressure $\mathcal P_l$ as long as the local shear stress $|\mathcal S_l|$ is large enough. Indeed, in Eq.~(10) the sign switching of the active pressure $\mathcal P_a$ (\emph{e.g.} $\mathcal P_a<0$ while $\mathcal P_l>0$) requires 
\begin{equation}
	\abs{\alpha}\mathcal S_l^2\gtrsim |\mathcal P_l|.
\end{equation} 
This sets the boundary between the regions with light shading and the regions with intermediate shading. Then, the extreme case where all active stress components switch sign (\emph{e.g.} $\mathcal P_a\pm \mathcal S_a<0$ while $\mathcal P_l\pm \mathcal S_l>0$) happens for 
\begin{equation}
	|\mathcal S_l|\lesssim|\mathcal P_l|\lesssim|\alpha|\mathcal S_l^2-|\mathcal S_l|, 
\end{equation}
\emph{i.e.} for $|\mathcal P_l|$ and $|\mathcal S_l|$ both larger than $2/|\alpha|$. This extreme case corresponds to the dark regions of Fig.~2(b).

\section{Finite-element simulations}
This section provides further details on the finite element simulations used to produce Fig.~3 of the main text.
Sec.~\ref{4a} describes our simulation methods. In Sec.~\ref{4b}, we discuss rectification in two additional fully nonlinear models.

	\subsection{Methods}\label{4a}
We solve the set of equations~\eqref{eq:system} via simulations with the finite element software Fenics~\cite{FenicsProject15} version 2019.2.0.dev0. We use a mesh with maximal size $l=0.01$ for $r_\text{in}=1$ and $r_\text{out}=2$, and another one with $l=0.1$ for $r_\text{out}=10$. They were both created with Gmsh version 4.4.1. In all figures, the error bars correspond to the differences between two meshes at $l$ and $l/10$, which gives roughly $5\%$ of $\mathcal S_l$ or $\mathcal P_l$ for all points. The meshes are created without enforcing rotational symmetry, which results in small non-zero values for the non-diagonal coefficients that should be zero in a continuum system (\emph{e.g.}, $\bar\sigma^l_{xy}$), as shown in Eq.~(8). However, we find that these values are smaller than $5\%$ of the diagonal coefficients in all simulations. In the geometry of Fig.~1, if we apply too large a deformation at $r_\text{in}$, we come into contact with the fixed boundary at $r_\text{out}$, which poses some numerical issues. Therefore, we can only perform accurate simulations up to about $\eta=\abs{e_0}+\abs{e_2}\sim0.6$.

	\subsection{Additional data}\label{4b}
    
\begin{figure}[!b]
\begin{minipage}[b]{.5\linewidth}
	\subfloat {\begin{tikzpicture}  
   		\node[anchor=north west,inner sep=0pt,outer sep=0pt] at (0,0) 
		{\includegraphics[width=.29\linewidth]{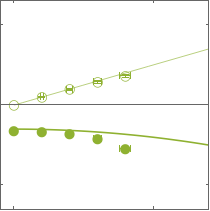}  };
    	\draw (-.06,-.25) node[right] {(a)};
    	\draw (-.85,.285) node[right] {$\mathcal P_l/\kappa=$\hspace{.6cm}$-0.3$};
    	\draw (-1.3,-2.3) node[right, rotate=90] {neo-Hookean};
    	\draw (-.75,-2.6) node[right, rotate=90] {$\kappa_1=\frac{1}{2},\,\mu_1=-\frac{3}{2}$};
    \end{tikzpicture}\hspace{-.2cm} 
	\begin{tikzpicture} 
    	\node[anchor=north west,inner sep=0pt,outer sep=0pt] at (0,0) 
   		{\includegraphics[width=.29\linewidth]{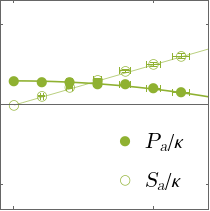}  };
    	\draw (-.06,-.25) node[right] {(b)};
    	\draw (.91,.285) node[right] {$0.3$};
    	\draw[gray] (1.6,-.77) node[right]{\scalebox{.9}{\color{white}0.5\hspace{1cm}}};
    	\draw[white, fill=white] (1.8,-1.53) rectangle (2.35,-2.35);
    	\draw[gray] (1.6,-1.75) node[right] {$\mathcal P_a/\kappa$};
    	\draw[gray] (1.6,-2.26) node[right] {$\mathcal S_a/\kappa$};
    	\draw[gray] (2.58,-.29) node[right]{\scalebox{.9}{1}};
    	\draw[gray] (2.58,-1.29) node[right]{\scalebox{.9}{0}};
    	\draw[gray] (2.53,-2.29) node[right]{\scalebox{.9}{-1}};
	\end{tikzpicture}  
		}  \vspace{-.445cm}\\ 
	\subfloat {\hspace{-.105cm}\begin{tikzpicture}
   		\node[anchor=north west,inner sep=0pt,outer sep=0pt] at (0,0) 
   		{\includegraphics[width=.29\linewidth]{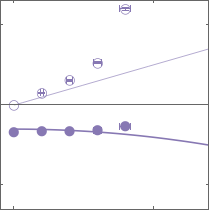}  };
   		\draw (-.06,-.25) node[right] {(c)};
   		\draw (-1.3,-2.5) node[right, rotate=90] {shear-stiffening};
   		\draw (-.75,-2.6) node[right, rotate=90] {$\kappa_1=\frac{1}{2},\,\mu_1=-\frac{3}{2}$};
   		\draw[gray] (-.024,-2.77) node[right]{\scalebox{.9}{0}};
   		\draw[gray] (1.5,-2.77) node[right]{\scalebox{.9}{0.5}};
   		\draw[gray] (1.,-3.19) node[right]{\scalebox{.9}{local shear stress $\mathcal S_l/\kappa$}};
   	\end{tikzpicture}\hspace{-1.66cm}
    \begin{tikzpicture}
   		\node[anchor=north west,inner sep=0pt,outer sep=0pt] at (0,0) 
   		{\includegraphics[width=.29\linewidth]{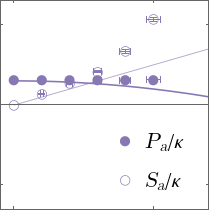}  };
   		\draw (-.06,-.25) node[right] {(d)};
   		\draw[white, fill=white] (1.8,-1.53) rectangle (2.35,-2.35);
    	\draw[gray] (1.6,-1.75) node[right] {$\mathcal P_a/\kappa$};
    	\draw[gray] (1.6,-2.26) node[right] {$\mathcal S_a/\kappa$};
    	\draw[gray] (2.58,-.29) node[right]{\scalebox{.9}{1}};
    	\draw[gray] (2.58,-1.29) node[right]{\scalebox{.9}{0}};
    	\draw[gray] (2.53,-2.29) node[right]{\scalebox{.9}{-1}};
   		\draw[gray] (-.024,-2.77) node[right]{\scalebox{.9}{0}};
   		\draw[gray] (1.5,-2.77) node[right]{\scalebox{.9}{0.5\hspace{1.12cm}}};
   		\draw[gray] (1.78,-3.19) node[right]{\color{white}\scalebox{.9}{$\mathcal S_l/\kappa$}};
    \end{tikzpicture}}    
\end{minipage}\hspace{-.95cm}
	\begin{tikzpicture}
		\draw (0,0) node[right] {$ $};
		\draw[gray] (0,4.4) node[right, rotate=-90] {\scalebox{.9}{active stresses}};
	\end{tikzpicture}
	\begin{tikzpicture} 
    	\draw (0,-.04) node[right, rotate=90] {$ $};
    	\draw (.5,2.4) node[right, rotate=90] {neo-Hookean};
    	\draw (0,6) node[above] {(e)};
    \end{tikzpicture}\hspace{-.9cm}
\begin{minipage}[b]{.5\linewidth}
	\subfloat {\begin{tikzpicture}  
   		\node[anchor=north west,inner sep=0pt,outer sep=0pt] at (0,0) 
		{\includegraphics[width=.29\linewidth]{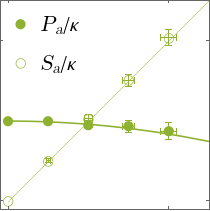}  };
    	\draw (-.85,.285) node[right] {$\cfrac{r_\text{out}}{r_\text{in}}=$\hspace{1.cm}2};
    	\draw[white, fill=white] (.4,-.1) rectangle (1.1,-1.);
    	\draw[gray] (.3,-.32) node[right] {$\mathcal P_a/\kappa$};
    	\draw[gray] (.3,-.8) node[right] {$\mathcal S_a/\kappa$};
    	\draw[white] (-.56,-.49) node[right]{\scalebox{.9}{0.2}};
    	\draw[white] (-.32,-2.48) node[right]{\scalebox{.9}{0}};
    \end{tikzpicture}\hspace{-.185cm}  
	\begin{tikzpicture} 
    	\node[anchor=north west,inner sep=0pt,outer sep=0pt] at (0,0) 
   		{\includegraphics[width=.29\linewidth]{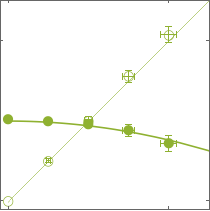}  };
    	\draw[white] (-.06,-.25) node[right] {(b)};
    	\draw (.95,.275) node[right] {$10$};
    	\draw (3.6,-1.285) node[right] {$\nu=0.1$};
    	\draw[white] (.32,-2.48) node[right]{\scalebox{.9}{0}};
   		\draw[gray] (2.56,-.49) node[right]{\scalebox{.9}{0.2}};
    	\draw[gray] (2.56,-2.48) node[right]{\scalebox{.9}{0}};
	\end{tikzpicture}  
		} \vspace{-.495cm}\\ 
	\subfloat {\hspace{-.cm}\begin{tikzpicture}
   		\node[anchor=north west,inner sep=0pt,outer sep=0pt] at (0,0) 
   		{\includegraphics[width=.29\linewidth]{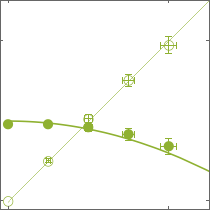}  };
   		\draw[white] (-.06,-.25) node[right] {(c)};
   		\draw[white] (-.85,-.585) node[right] {$ $};
   		\draw[gray] (-.075,-2.76) node[right]{\scalebox{.9}{0}};
   		\draw[gray] (1.81,-2.76) node[right]{\scalebox{.9}{0.2}};
   		\draw[gray] (1.04,-3.19) node[right]{\scalebox{.9}{local shear stress $\mathcal S_l/\kappa$}};
   		\draw (.5,-3.44) node[right, rotate=90] {\color{white}a};
   	\end{tikzpicture}\hspace{-1.71cm}
   	\begin{tikzpicture}
   		\node[anchor=north west,inner sep=0pt,outer sep=0pt] at (0,0) 
   		{\includegraphics[width=.29\linewidth]{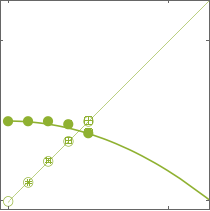}  };
   		\draw[white] (-.06,-.25) node[right] {(d)};
   		\draw (3.6,-1.285) node[right] {$\nu=0.8$};
   		\draw[white] (.32,-2.48) node[right]{\scalebox{.9}{0}};
   		\draw[gray] (2.56,-.49) node[right]{\scalebox{.9}{0.2}};
    	\draw[gray] (2.56,-2.48) node[right]{\scalebox{.9}{0}};
   		\draw[gray] (-.075,-2.76) node[right]{\scalebox{.9}{0}};
   		\draw[gray] (1.81,-2.76) node[right]{\scalebox{.9}{0.2}};
   		\draw[gray] (1.94,-3.1) node[right]{\scalebox{.9}{\color{white}$\mathcal S_l/\kappa$}};
   		\draw (1.5,-3.44) node[right, rotate=90] {\color{white}a};
    \end{tikzpicture} }
\end{minipage}\hspace{-2.5cm}
	\begin{tikzpicture}  
    	\draw (0,-.04) node[right, rotate=90] {$ $};
    	\draw[gray] (.5,4.4) node[right, rotate=-90] {\scalebox{.9}{active stresses}};
    \end{tikzpicture}\hspace{1.4cm}
   \caption{Additional plots of the coarse-grained stresses are in agreement with Eq.~(10) up to intermediate stresses and in a large scale of parameters. (a-b)~Rubber-like neo-Hookean model [$a=b=0$ in Eq.~(11)]. (c-d)~Fiber-like shear-stiffening model of Eq.~\eqref{eq:E_fibrelike} with $c=10$, following the predictions up to the point where stresses diverge (when $\eta\sim1/\sqrt c$). For all plots, $\nu=0.1$ and $r_\text{out}/r_\text{in}=2$. (e)~For $\mathcal P_l=0.1\kappa$, the predictions at $\nu=0.8$ and $r_\text{out}/r_\text{in}=10$ remain quantitatively accurate.}
\label{FES_SI1}
\end{figure}

\begin{figure}[!t]
	\begin{tikzpicture}  
    	\draw (0,0) node[right, rotate=90] {\color{white}a};
    	\draw (0,2.35) node[right, rotate=90] {neo-Hookean};
    	\draw[gray] (.7,1.95) node[right, rotate=90] {\scalebox{.9}{active pressure $\mathcal P_a/\kappa$}};
    \end{tikzpicture}\hspace{-1.cm}
\begin{minipage}[b]{.5\linewidth}
	\centering
	\subfloat {\begin{tikzpicture}  
   		\node[anchor=north west,inner sep=0pt,outer sep=0pt] at (0,0) 
		{\includegraphics[width=.29\linewidth]{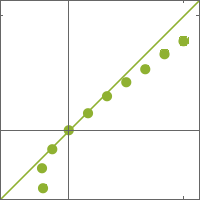}  };
    	\draw (-.06,-.25) node[right] {(a)};
    	\draw (-.85,.485) node[right] {$\cfrac{r_\text{out}}{r_\text{in}}=$\hspace{1.cm}2};
    	\draw[gray] (-.31,-.2) node[right]{\scalebox{.9}{3}};
    	\draw[gray] (-.31,-1.7) node[right]{\scalebox{.9}{0}};
    \end{tikzpicture}\hspace{-.195cm}  
	\begin{tikzpicture} 
    	\node[anchor=north west,inner sep=0pt,outer sep=0pt] at (0,0) 
   		{\includegraphics[width=.29\linewidth]{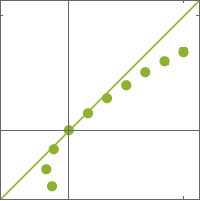}  };
    	\draw (-.06,-.25) node[right] {(b)};
    	\draw (.95,.475) node[right] {$10$};
    	\draw (2.7,-1.285) node[right] {$\nu=0.1$};
	\end{tikzpicture}  
		} \vspace{-.43cm}\\ 
	\subfloat {\hspace{-.cm}\begin{tikzpicture}
   		\node[anchor=north west,inner sep=0pt,outer sep=0pt] at (0,0) 
   		{\includegraphics[width=.29\linewidth]{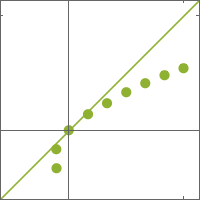}  };
   		\draw (-.06,-.25) node[right] {(c)};
   		\draw (-.85,-.585) node[right] {$ $};
   		\draw[gray] (-.31,-.2) node[right]{\scalebox{.9}{3}};
    	\draw[gray] (-.31,-1.7) node[right]{\scalebox{.9}{0}};
    	\draw[gray] (-.5,-1.7) node[right, rotate=90]{\scalebox{.9}{$ $}};
   		\draw[gray] (.705,-2.76) node[right]{\scalebox{.9}{0}};
   		\draw[gray] (2.2,-2.76) node[right]{\scalebox{.9}{3}};
   		\draw[gray] (1.24,-3.1) node[right]{\scalebox{.9}{local pressure $\mathcal P_l/\kappa$}};
   	\end{tikzpicture}\hspace{-1.495cm}
   	\begin{tikzpicture}
   		\node[anchor=north west,inner sep=0pt,outer sep=0pt] at (0,0) 
   		{\includegraphics[width=.29\linewidth]{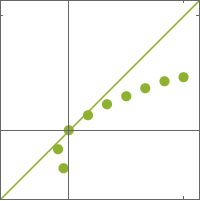}  };
   		\draw (-.06,-.25) node[right] {(d)};
   		\draw (2.7,-1.285) node[right] {$\nu=0.8$};
   		\draw[gray] (.705,-2.76) node[right]{\scalebox{.9}{0}};
   		\draw[gray] (2.2,-2.76) node[right]{\scalebox{.9}{3}};
   		\draw[gray] (.94,-3.1) node[right]{\scalebox{.9}{\color{white}$\mathcal P_l/\kappa$}};
    \end{tikzpicture} }
\end{minipage}
	\begin{tikzpicture}
   		\node[anchor=north west,inner sep=0pt,outer sep=0pt] at (0,0) 
		{\includegraphics[width=.181\linewidth]{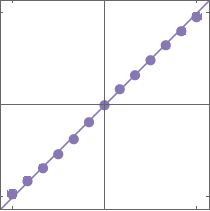}  };
		\draw (0,-4.95) node[right, rotate=90] {\color{white}a};
		\draw (-.06,-.25) node[right] {(e)};
		\draw (.1,.5) node[right] {$\cfrac{r_\text{out}}{r_\text{in}}=2,\ \ \nu=0.1$};
    	\draw (.3,-.75) node[right] {$c=10$};
    	\draw (-.95,-2.8) node[right, rotate=90] {shear-stiffening};
    	\draw[gray] (-.31,-.19) node[right]{\scalebox{.9}{3}};
    	\draw[gray] (-.31,-1.61) node[right]{\scalebox{.9}{0}};
    	\draw[gray] (-.4,-3.04) node[right]{\scalebox{.9}{-3}};
    	\draw[gray] (-.02,-3.39) node[right]{\scalebox{.9}{-3}};
    	\draw[gray] (1.435,-3.39) node[right]{\scalebox{.9}{0}};
   		\draw[gray] (2.85,-3.39) node[right]{\scalebox{.9}{3}};
    \end{tikzpicture}
   \caption{Investigation of the $\mathcal P_a\sim \mathcal P_l$ relationship of Eq.~(10) for $\mathcal S_l=0$. The predictions at zero shear stress (\emph{i.e.} $|\mathcal S_l|,|\mathcal S_a|<0.01\kappa$) remain reasonably accurate up to a few $\mathcal P_l/\kappa$. (a-d) In the neo-Hookean case, the nonlinear terms are such that $\mathcal P_a<\mathcal P_l$ and become increasingly significant as $\nu$ and $r_\text{out}/r_\text{in}$ increase. Specifically, the agreement deteriorates from panel (a) to panels (b) and (c), to panel (d). (e) In the shear-stiffening case these nonlinear terms are first decreased up to $c\sim2$-5 [see Eq.~\eqref{eq:E_fibrelike}] and then increased in the opposite direction (such that $\mathcal P_a>\mathcal P_l$).} 
\label{FES_SI2}
\end{figure}

In the main text, we studied two models with clear buckling and anti-buckling behaviors, which lead to a readily observable reversal of the active pressure sign, due to rectification. Here, we study two other models where these behaviors are less pronounced: a standard neo-Hookean model of rubber, and another one which can mimic the shear-stiffening behavior of fiber networks. Consistent with analytical predictions, these systems display a smaller propensity for rectification, and we show that the predictions of Eq.~(10) remain valid up to intermediate stress values.

In the fully nonlinear model with the elastic energy density of Eq.~(11), we introduced the parameters $a,b$ such that $\kappa_1=1/2-3a$ and $\mu_1=-3/2-b$. In Fig.~3, we showed the good agreement between the weakly nonlinear predictions of Eq.~(10) and finite-element simulations for the bucklable and anti-bucklable models obtained by setting $(a,b)=(-1/6, -5/2)$ and $(3/2, 5/2)$ to obtain $(\kappa_1,\mu_1)=(1,1)$ and $(-4,-4)$ respectively. We now consider the neo-Hookean model, obtained by setting $a=b=0$ in Eq.~(11). It has $\kappa_1=1/2$, $\mu_1=-3/2$ and we recover the predicted small tendency to rectify towards contraction, see Fig.~\ref{FES_SI1}(a,b).

\newpage

We then investigate a variant of the neo-Hookean model which has the shear-stiffening behavior $G\propto \sigma_{xy}^{3/2}$ characteristic of fiber networks under large strains. Its elastic energy density reads~\cite{shokef_scaling_2012, knowles_77}
\begin{equation}\label{eq:E_fibrelike}
	E=\frac{\kappa}{2}\left(J-1\right)^2 + \frac{\mu}{2c}\Big[1-c\big(I/J-2\big)\Big]^{-1},
\end{equation}
where $J=\det(\mathbf1+\bm\eta)$ and $I=\Tr(\mathbf1+2\bm\varepsilon)$. Here, the shear strain threshold for the stiffening behavior corresponds to a fraction of $1/\sqrt c$, the strain at which the shear stress diverges. This is such that the neo-Hookean model is recovered for $c=0$. This model still has $\kappa_1=1/2$, $\mu_1=-3/2$, which corresponds to the same tendency to rectify towards contraction as in the neo-Hookean case. Indeed, $c$ only affects higher order nonlinearities. As shown in Fig.~\ref{FES_SI1}(c,d) where $c=10$, we recover Eq.~(10) at small stress. But due to the shear stress divergence at finite shear strain in Eq.~\eqref{eq:E_fibrelike}, our predictions fail when $\mathcal S_l/\kappa\gtrsim1/\sqrt{c}$. Furthermore, the agreement between the simulations with the neo-Hookean model and Eq.~(10) remains quantitative in the small stress regime for all considered values of the Poisson's ratio $\nu$ and the ratio of the boundary radius to the active unit radius $r_\text{out}/r_\text{in}$, see Fig.~\ref{FES_SI1}(e). 

We finally compare the dependence of $\mathcal P_a$ with $\mathcal P_l$ at zero shear stress with the weakly nonlinear prediction $\mathcal P_a\sim \mathcal P_l$. As displayed in Fig.~\ref{FES_SI2}, we recover the prediction for small stresses, but higher order nonlinearities induce significant deviations for $\mathcal P_l/\kappa$ outside of $[-0.3,1]$. In the end, we see that the predictions of Eq.~(10) relating the active and local stress components fail when either the local pressure or the local shear stress become comparable to the bulk modulus $\kappa$. Our weakly nonlinear predictions additionally fail close to stress divergences, \emph{i.e.} when the nonlinearities become too significant compared to the linear terms.